\documentclass[conference]{IEEEtran}
\IEEEoverridecommandlockouts
\usepackage{cite}
\usepackage{amsmath,amssymb,amsfonts}
\usepackage{algorithmic}
\usepackage{graphicx}
\usepackage{textcomp}
\usepackage{xcolor}
\usepackage{comment}
\usepackage{mathtools}
\usepackage{xspace}
\usepackage{comment} 
\usepackage{tikz}

\newcommand\copyrighttext{%
  \footnotesize This work has been accept  to the IEEE LATINCOM2020. Copyright 978-1-7281-8903-1/20/\$31.00~\copyright2020 IEEE.}
\newcommand\copyrightnotice{%
\begin{tikzpicture}[remember picture,overlay]
\node[anchor=south,yshift=8pt] at (current page.south) {\fbox{\parbox{\dimexpr\textwidth-\fboxsep-\fboxrule\relax}{\copyrighttext}}};
\end{tikzpicture}%
}

\newcommand{\sep}{\hspace{8 mm}}
\def\BibTeX{{\rm B\kern-.05em{\sc i\kern-.025em b}\kern-.08em
\kern-.1667em\lower.7ex\hbox{E}\kern-.125emX}}

\definecolor{color1}{RGB}{198, 90, 103}
\definecolor{color2}{RGB}{175, 42, 48}
\definecolor{color3}{RGB}{0,128,128}
\definecolor{color4}{RGB}{179, 43, 59}
\definecolor{color5}{RGB}{128,0,128} 
\definecolor{color6}{RGB}{4,159,255} 

\newcommand{\cp}[1]{\textcolor{color3}{#1}}

\usepackage[english,ruled,noline,linesnumbered,noend]{algorithm2e}

\title{
Detecting FDI Attack on Dense IoT Network with Distributed Filtering Collaboration and Consensus
\\
}

\author{
\IEEEauthorblockN{{\bf Carlos Pedroso Junior\IEEEauthorrefmark{1}},  
{\bf Aldri Santos\IEEEauthorrefmark{1}}, 
{\bf Michele Nogueira\IEEEauthorrefmark{1}}}

\IEEEauthorblockA{\IEEEauthorrefmark{1}Wireless and Advanced Networks Laboratory (NR2) - UFPR, Brazil  \\
Emails: \{capjunior, aldri, michele\}@inf.ufpr.br 
}}

\IEEEaftertitletext{\vspace{-1.0\baselineskip}}

\begin{document}
\maketitle
\copyrightnotice
\thispagestyle{plain} \pagestyle{plain} 

\begin{abstract}
The rise of IoT has made possible the development of 
personalized services, like industrial services  that often deal with massive amounts of data.  However, as IoT grows, its threats are even greater. The false data injection  (FDI) attack stands out as being one of the most harmful to data networks like IoT. The majority of current systems to handle this attack do not take into account the data validation, especially on the data clustering service. This work introduces 
CONFINIT, an intrusion detection system against FDI attacks on the data dissemination service into dense IoT. It combines watchdog surveillance and collaborative consensus among IoT devices for getting the swift detection of attackers. CONFINIT was evaluated in the NS-3 simulator into a dense industrial IoT and it 
has gotten detection rates of 99\%, 3.2\% of  false negative and 3.6\% of false positive rates, adding up to 35\% in 
clustering without FDI attackers.
\end{abstract}


\section{Introduction}
Internet of Things (IoT) enables the connection of a variety of physical devices through technologies like RFID, GPS and NFC, among others. IoT objects usually hold features like identity, physical attributes, mobility level, and many of them 
through smart interface establish communication with each other~\cite{gubbi2013internet}. IoT is essential to gather, disseminate, and 
arrange the data volume required by  various  applications to make its decisions~\cite{minoli2017iot}. For instance, the Industrial Internet of Things (IIoT) has recently brought attention for
enabling diverse industrial devices to play in an organized and synchronized way. Though, IoT still exhibits various vulnerabilities and challenges to achieve better robustness~\cite{mumtaz2017massive} due to its unique aspects. 
The  massive amount of data due to the interaction among multiple devices exposes dense IoTs to 
data security threats. 
Those 
environments typically rely on mobile and fixed devices, and the infrastructure can change through ties among devices~\cite{gubbi2013internet}. 
Thus, IoT has been target of attacks
that violate many of the security attributes, like integrity, authenticity, and availability of services; including the data dissemination,
which hinders the performance 
of various~applications~\cite{mendez2018internet}.

Among the internal threats to the IoT data dissemination service, False Data Injection (FDI) Attack stand out as one of the most harmful due to the inconsistent data 
yielded by them 
and the unpredictability of its actions~\cite{yang2017robust}.
Due to that unpredictability, the  attack detection is difficult, as 
captured devices are usually authenticated on networks and perform their standard data collection and dissemination functions~\cite{deng2016false}. Furthermore, attacks can carry out continuously and at distinct times, disturbing the network. The FDI attack take place when a device take control 
others, and then alter, fabricate, or manipulate its data, or when the device itself behaves in misconducting. 
Thus, it is essential quickly to identify and isolate malicious devices on dense IoT networks to avoid long malfunction times 
and data inconsistencies.

Although there are  approaches to handle FDI attacks in WSNs~\cite{lu2012becan}, Smart Grids~\cite{li2017distributed} or even IoT~\cite{yang2017robust}, they have usually failed or are not suited to dense IoTs because they do not analyze the data and few employ  collaborative detection. Additionally, systems, like En-Route Filtering Scheme, Collaborative Detection and Intrusion Detection Systems (IDS), have been commonly intended to monitor anomalies based on network traffic and not to the gathered data~\cite{raptis2019data}.
IDS are a robust approach against 
attacks in diverse IoT contexts~\cite{yang2017robust}. 
But, 
when IDS are not correctly employed in a IoT, they can bring up new vulnerabilities.~Therefore, it is mandatory to detect and isolate IoT threats in a distributed manner  to assure greater robustness for the service of data dissemination. An effective manner 
to face 
FDI attacks consists of~usage jointly of collaborative consensus~\cite{colistra2014task} for decision making~among network~devices and the watchdog monitoring~\cite{yang2017robust}, since their techniques would allow us swiftly 
the detection and identification  of nodes that exhibit FDI attack behavior over the time.

This work presents CONFINIT (\emph{\textbf{CON}sensus Based Data \textbf{FI}lteri\textbf{N}g for \textbf{I}o\textbf{T}}) a system to mitigate and isolate false data injection attacks on data dissemination services in dense IoT networks. CONFINIT aims to detect and isolate 
data misbehaving devices  on the clustering service. For that, CONFINIT applies clustering by data similarity to handle the density of devices~on the network. It also combines a watchdog strategy between devices for 
self-monitoring as well as a collaborative consensus for decision making on FDI misbehaving devices. 
Simulations showed that CONFINIT achieved attack detection rates of 99\% 
in  
dense IIoT scenarios, only up to 3.2\% of false negatives and 
3.6\% of false positives, increasing up to 35\% the number of clustering without~attackers. 

This paper is organized as follows: Section~\ref{sec:rel} 
presents 
the related work. Section~\ref{sec:sys} defines the model and assumptions taken by CONFINIT. Section~\ref{sec:CONF}  describes 
the CONFINIT components 
and their operation. Section~\ref{sec:ana}  shows
an evaluation of CONFINIT
and 
the results obtained. Section~\ref{sec:con} presents conclusions and future work.

\section{Related Work}
\label{sec:rel}



In~\cite{yang2017robust}, an IDS based on anomaly detection applies watchdog to mitigate False Data Injection attacks. They attempt 
to predict natural events by using environmental IoT surveillance data through Hierarchical Bayesian Time-Space (HBT) monitoring and a statistical decision strategy in a sequential probability test supports to identify malicious activities. However, the HBT model is energetically costly and the probability test, although effective, overloads the network and negligence the data validation. 
In~\cite{santos2019clustering}, an IDS to mitigate sinkhole and selective forwarding attacks in dense IoT arranges the network in clusters and classifies the nodes in categories to get better communication. Besides that, a watchdog strategy in levels monitors the relationship between received and transmitted data to determine nodes with malicious behavior. However, computing trust between nodes only by a difference of data received and transmitted makes it hard to identify nodes that manipulate the collected data.
In~\cite{lu2012becan}, a cooperative authentication scheme aimed at filtering false data in WSNs, so that all nodes require a fixed number of neighbors to manage the authentication in a distributed 
across data routing to the base station. The scheme also adopts the compression bit technique aiming to avoid overloading the communication channel, making it appropriated to filter injected data since the authentication happens point-to-point. However, this scheme 
is fragile to face the data manipulation and does not identify compromised devices.
In~\cite{yu2010dynamic}, a scheme of dynamic en-route filtering to handle false data and DoS attacks on WSNs employs a group of authentication keys by node that are used to endorse reports. Authentication is guaranteed by a group of nodes chosen before the network starts. Each node offers its keys to the forwarding nodes which in turn should propagate the received keys, allowing the forwarders to verify all the reports. However, the scheme also does not take into account data manipulation and causes an additional cost to the network due to the constant exchange of keys between devices. 



In~\cite{toulouse2015consensus}, a distributed system to detect anomalies caused by DDoS attacks applies an averaged consensus protocol among the participants. It analyzes each data collection point using a Bayes classifier. The analysis occurs redundantly, parallel to the level of each data collection point, which avoids the single point of failure. The distributed consensus favors collaborative decision making. Though, the communication cost  among the participants overloads the network and diminish the system’s effectiveness. In~\cite{kailkhura2015consensus}, a distributed consensus system using distributed weighted average makes use of an algorithm to allow adaptation to local rules stipulated by the network, where a learning technique estimates operating parameters or weight of each node to automate local merge or update rules to mitigate attacks. The algorithm acts in a distributed manner, but it does not take the interaction among network devices as an impact factor, which allows malicious actions.

\section{Environment 
Model}
\label{sec:sys}
\begin{figure}[h] 
\centering
\hspace{-0.4cm}
\includegraphics[width=0.85\linewidth]{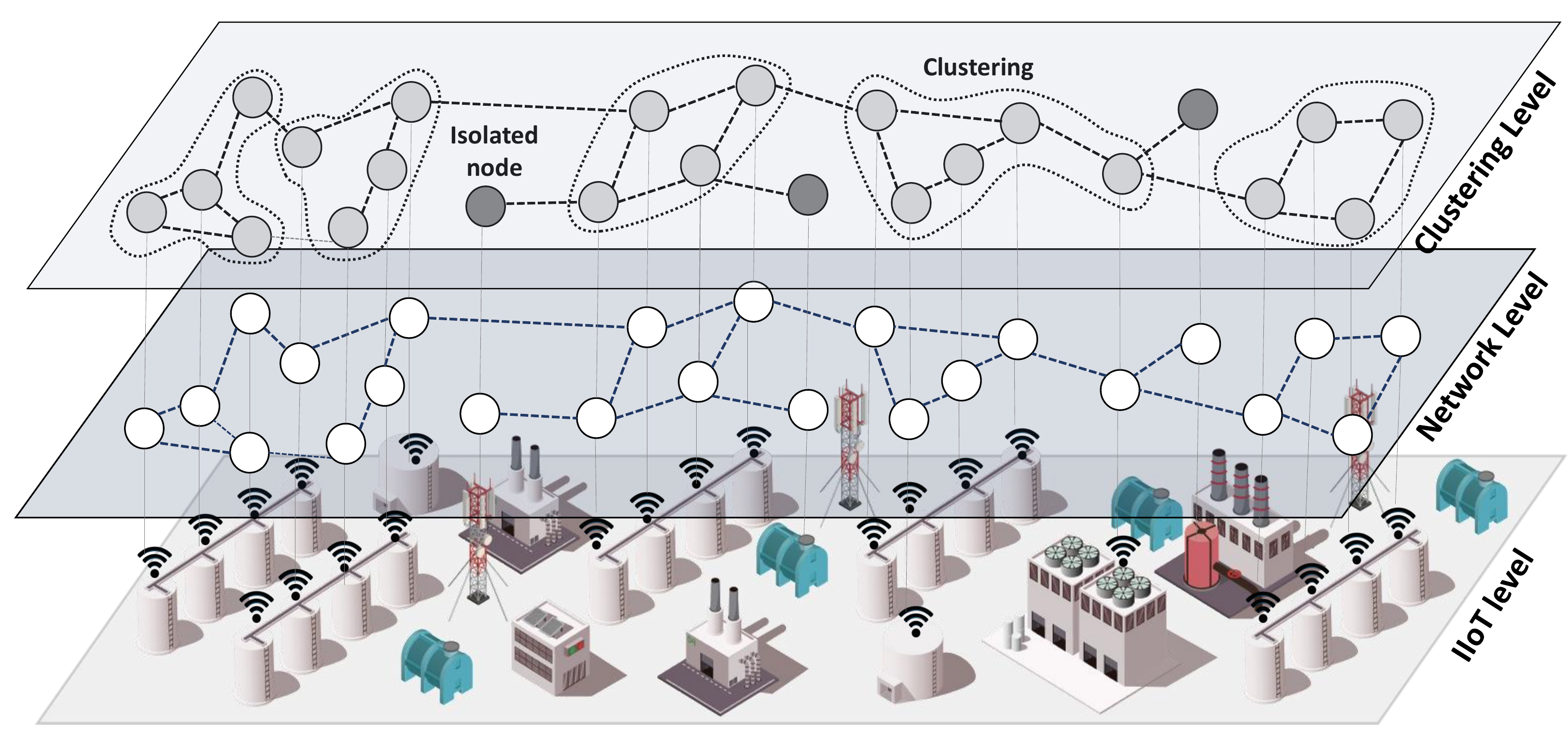}
\caption{Network model}
\label{Fig:net}
\end{figure}

This section describes our assumption on the network infrastructure and the FDI attack model. 
The network model consists of a three-level structure, whose 
the first one means the IIoT objects, the second one carries out the
communication 
among objects, and the third one coordinates the virtual clustering of the objects.
Fig.~\ref{Fig:net} illustrates all 
levels and their relationship.

\subsubsection*{\textbf{Network model}}
A massive IoT network composed by 
a set of objects (nodes), 
as~an industrial environment, 
denoted by 
$N = \{n_1, n_2, n_3,..., n_n \}$, where $n _i\in N$. Each node $n_i$ has a unique physical address $Id$ that identifies it on the network. 
Each node clustering is 
a subset ${A_{\alpha}}\subseteq N$, where each node $n_{i}\in{N}$. The node communication carries out in the wireless medium with an asynchronous channel and subject to the packet loss.
In addition, all nodes work with the same transmission range to establish clustering. 
We consider that nodes do not suffer from energy restriction in the context of an 
IIoT network. 
%
{\it \textbf{FDI Attack model:}}
The FDI attack takes place in two 
manners: {\it i)}  
through the capture of the object and manipulation of the data by either altering or falsifying it;
{\it ii)}  
the object itself is the attacker, and it changes, fabricates, or manipulates its own data. The attacker exploits vulnerabilities arising from other attacks, in addition to the attacker 
with full knowledge about the network and works coordinated~\cite{yang2017robust, yaqoob2017rise}.

\section{CONFINIT Architecture}
\label{sec:CONF}
\begin{figure}[ht]
 \centering 
 \includegraphics[width=1\linewidth]{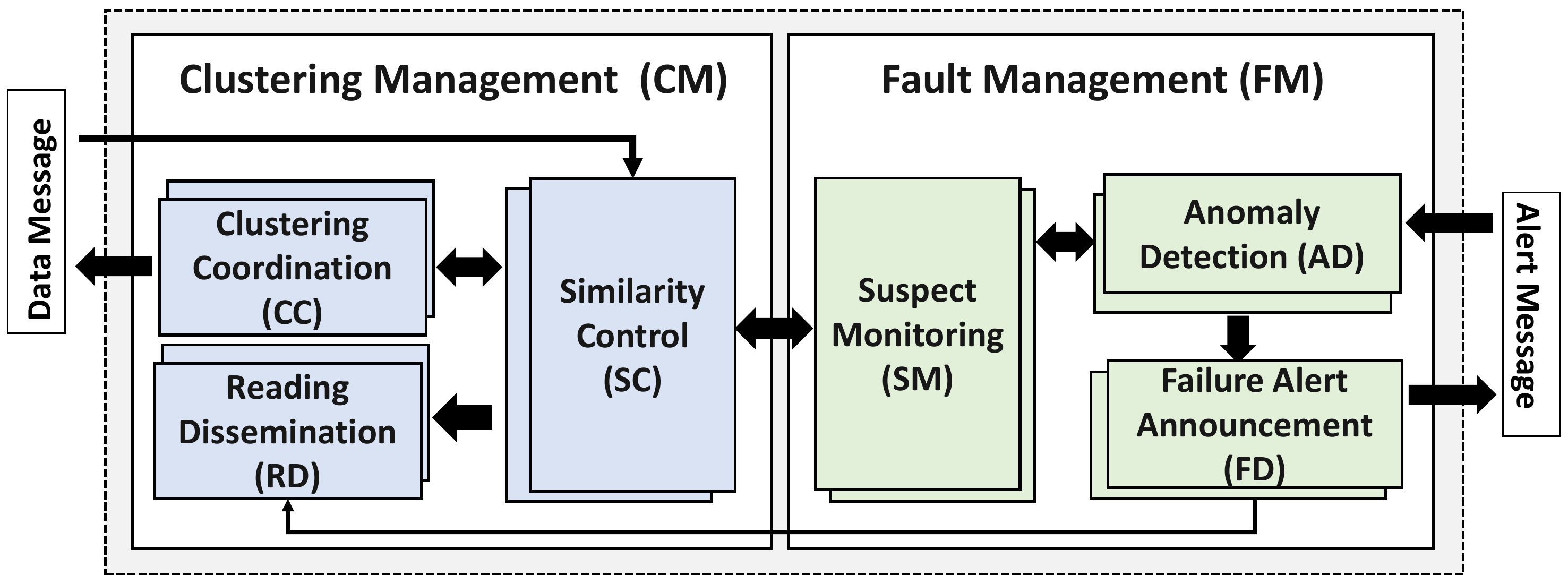}
 \caption{The CONFINIT architecture}
 \label{Fig:arqu}
\end{figure}

The CONFINIT architecture consists of the \textbf{Clustering Management} and \textbf{Fault Management} modules that work jointly to ensure the secure dissemination of data in the network, as 
shown in Fig~\ref{Fig:arqu}. The former 
arranges
the devices 
in clusters and
the latter deals with
the monitoring of devices,
and 
detection and isolation of FDI misbehaving devices.

\textbf{The Clustering Management (CM) Module} ensures the formation and maintenance of clusters of devices in the network. ~\textbf{CM} builds the clusters based on the similarity threshold of readings from nearby devices (nodes) and establishes 
the integration of nodes 
in a cluster. 
Thus, when receiving a data message 
~\textbf{CM} analyzes 
the identification, the number of neighbors and the readings of those neighbors. \textbf{CM} consists of
the \textit{Similarity Control (SC)},~\textit{Clustering Coordination (CC)} and~\textit{Reading Dissemination (RD)} components. $SC$ acts by receiving and interpreting messages exchanged among the nodes.
$CC$ takes care to form and maintain clusters by using the similarity of readings among nodes, being also responsible for electing leader.  $RD$ acts for disclosing the device's reading, number of readings and neighbors,~so that all nodes that receive the message will know when they~are~part of a cluster.

\textbf{The Fault Management (FM) Module} ensures the security of data dissemination among 
nodes so that only authentic readings are disseminated over the network. For that, it takes 
the \textit{Suspect Monitoring 
(SM)},~\textit{Anomaly Detection (AD)}, and~\textit{Failure Alert 
Announcement (FA)} components. $SM$ monitors other nodes that do not respect the similarity threshold. 
$AD$ deals with the collaborative consensus and the standard deviation technique to detect 
FDI misbehaving nodes.
The collaborative consensus is the 
agreement and uniformity of opinions that nodes establish through the exchange of information 
among their cluster neighbors. The usage of standard deviation aims to determine how many discrepant the readings are in a given moment. Lastly, 
$FA$ acts to isolate FDI attacking nodes and alert 
other members about the threat. Thus, when detecting a FDI misbehaving node, the detector node propagates an alert message to the cluster leader that spreads it across the network.
The alert message carries the attacker $Id$ and its individual reading as well as the detector node $Id$. The sending time of the alert message takes into account the exact timing of detection and propagation.

\subsection{Cluster configuration}
~\textbf{CM} arranges the network into clusters, based on leaders, to create a network topology. Initially, nodes begin their operations in an isolation way, transmitting and collecting exchanged data messages from neighboring nodes for the cluster formation. Algorithm~\ref{alg:clus} presents how the clustering 
works. Similarity clustering prevents data redundancy and ensures that a single read represents data from all cluster nodes.
Periodically, after a data gathering, each node broadcasts one data message, with the following information $<Id$, ${L_{ind}}$, ${L_{agr}}$, ${N_ {viz}}>$, 
for mentioning its identifier $Id$, its current reading ${L_{ind}}$, the aggregate reading average of nodes with similar readings on its neighborhood ${L_{agr}}$ and the amount of neighbors ${N_{viz}}$ (\textit{l}.1-\textit{l}.5). The  $WaitInterval$ function establishes a random infimum delay to avoid simultaneous transmissions, and thus collision of messages among the nodes (\textit{l}.3).
Upon receiving a data message (\textit{l}.6), the node will know the source $org$, the individual reading $iR$ from the issuing node, the aggregate reading $aR$ from its neighborhood, and the amount 
of neighboring nodes by 
$nR$. Initially, the node updates the received information (\textit {l}.7), so that average aggregate readings enable
to find out
whether 
the current node reading satisfies the similarity threshold (\textit {l}.8). Thus, the similarity check up is made 
in (\textit {l}.9). The ${N _ {viz}}$ structure neighborhood is updated by setting 
adding 
$org$ when the similarity threshold is satisfied  (\textit{l}.10) or removing when the node not satisfied (\textit{l}.12). This process 
occurs dynamically on each network node, ensuring that everyone can keep their neighborhood structure up to date.

\begin{algorithm}[ht]
\relsize{-1.9}

\SetAlgoLined
\textbf{procedure} \textsc{SendDataMessage}() \\
\sep  $Send(Id, {L_{ind}}, {L _{agr}},
|{N _{viz}}|)$ \\
\sep $WaitIntervalnextmsg$ \\
\sep $RControlTimerExpire()$\\
\textbf{end procedure}\\

\BlankLine

\textbf{procedure} \textsc{ReceiveDataMessage}{($Org,iR,aR,nR$)} \\
\sep ${N _{viz}}[Org] \leftarrow \{iR, aR, nR\}$ \\
\sep $localRead \leftarrow {L _{agr}}()$ \\
\sep \textbf{if} ${(|iR - localRead| < Threshold)}$ \textbf{and}  ${(|{L _{ind}}() - aR| < Threshold)}$ \textbf{then} \\
\sep \sep ${N _{viz}} \leftarrow S{N _{viz}} \bigcup \{org\}$\\
\sep \textbf{else if} \\
\sep \sep ${N _{viz}} \leftarrow S{N _{viz}} \bigcap \{org\}$\\
\sep \textbf{end if}\\
\textbf{end procedure}\\
\caption{Clustering establishment}
\label{alg:clus}
\end{algorithm}

{\scriptsize 
\begin{equation}
\label{eq:firefc}
	\left | Y -
	\frac{X + \sum_{v \in S{N_{viz}}} ({N_{viz}}[v].aR * {N _{viz}}R[v].nR) }
	{1 + \sum_{v \in S{N _{viz}}} ({N_{viz}}[v].nR)}
 \right |< CThresh
\end{equation}
} 

The computation of the similarity relationship between two nodes makes use of 
their reading values and  the threshold value between them. Equation~\ref{eq:firefc} verifies the similarity between two readings, following the DDFC model developed by~\cite{gielow2015data}, whose  model enables adaptability and accuracy to dense environments like IIoT. The equation considers the node readings, the amount of neighbors, and the aggregate readings of those neighbors. $X$ means the current reading of the node and $Y$ the reading with which it is being compared. The comparison outcome  determines which  readings respect the similarity threshold ($CThresh$), and thus the node can  join  the same~cluster. In the Dynamic Data-aware Firefly-based Clustering (DDFC) model, the value of the similarity threshold is strictly dependent on the application and the data use~\cite{gielow2015data}.

\subsection{Fault detection}
The nodes whose forwarded values do not respect the threshold of similarity of readings 
over 
the clustering formation stage,
firstly, are 
classified 
as  suspects and compose the individual suspect list of the node whom they 
form a cluster.
The usage of the suspect list enables 
a better evaluation of fault detection, since devices may at times present faults behavior, which does not characterize attacker behavior. Algorithm~\ref{alg:dect} details the operation of the fault detection module within the network against FDI threats. 
the detection action takes effect effectively after the first message exchange, as nodes need other messages to compare. Thus, at first, \textbf{FM} analyzes if a given node
belongs to the suspect list (\textit{l}.1-\textit{l}.7).
In the case of not being a suspect,
but its readings are 
suspicious the module inserts the node on the suspect list (\textit{l}.8-\textit{l}.16). Otherwise, if it belongs to the list, buts its readings meet the defined threshold, the module removes it from the list (\textit{l}.17-\textit{l}.20).


\begin{algorithm}[ht] 
\relsize{-1.9}
\textbf{procedure} \textsc{CheckSuspicious}  $(Id, ConsensusParticpant)$ \\
\sep \textbf{if}$(ID$ $\in$  $SuspectList$ $\And$ $ConsensusParticpant$ == $False$)\\
\sep \sep $return 1$ \\
\sep \textbf{else if}$(ID$ $\in$  $SuspectList$ $\And$ $ConsensusParticpant$ == $True$)\\
\sep \sep $return 2$ \\
\sep \textbf{end if}\\
\textbf{end procedure}\\

\BlankLine
\textbf{procedure} \textsc{CheckSuspicious}$  (Id, Read)$\\
\sep $Valid \longleftarrow checkSuspicious  (Id, ConsensusParticpant)$ \\
\sep \textbf{if}$(Read \leq Thresholdconsensus$)\\
\sep \sep $Switch \longleftarrow Valid$\\
\sep \sep \textbf{\textit{Case 1}}\\
\sep \sep  $Atklist \leftarrow  Ataklist \bigcup \{Id, Read\}$ \\
\sep \sep  \textbf{\textit{Case 2}}\\
\sep \sep  $Suspectlist \leftarrow Suspectlist \bigcap\{Id, Read\}$ \\
\sep \sep  $Switch \longleftarrow Valid$\\
\sep \textbf{else}\\
\sep \sep $Suspectlist \leftarrow Suspectlist \bigcup\{Id, Read\}$ \\
\sep \textbf{end if}\\
\textbf{end procedure}\\

\BlankLine
\caption{Attacker detection}
\label{alg:dect}
\end{algorithm}

Equation~\ref{eq:conse} determines the consensus computation to establish the deviation of the measured values. For this purpose, the readings data collected from the consensus participants are used to compare them. So we use a data set~\textit{D} = $({{d_{i}},{d_{i + 1}}},...,{{d_{n}})}$, which represents the data samples to check. The consensus calculation is indicated by $\sum_{i=1}^{n}$, which adds up all values of the~\textit {D} set, from the first position~\textit{(i=1)} to the  position~\textit{d} $\in$ \textit{D}. The value of $d_{i}$ represents at the~\textit{i} position in the data set~\textit{D}. $M_{A}$ represents the arithmetic mean of the data.~\textit{N} represents the amount of data to be evaluated in forming consensus. The~\textit{Thresholdconsensus} means the default threshold and it can change according to the type of the data evaluated and the type of application. 


{\scriptsize 
\begin{equation}
\label{eq:conse}
DP =
\sqrt{\frac{\sum_{i=1}^{n}\left (d_{i} - M_{A}  \right )^{2}}{N}}
\leq Threshold consensus
\end{equation}
} 

\subsection{Operation}
\begin{figure}[ht]
\centering    
\includegraphics[width=1\linewidth]{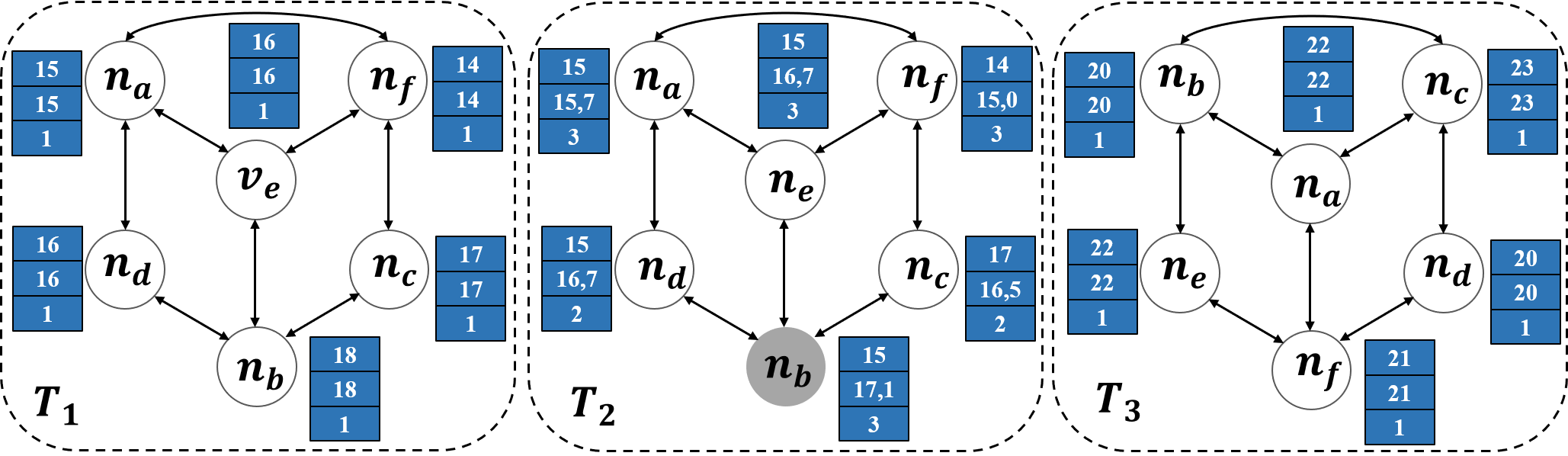}
\caption{Clustering formation over the time}
\label{Fig:funci}
\end{figure}

~\cp{The clustering formation stage
occurs dynamically for each node of the IoT network.} Interactions among nodes take place under space and time quantities, 
in this way
messages are sent and received by nodes into the transmitters transmission range. Each node broadcasts one data message with
its reading, its neighbors readings, and the number of neighbors.
The receiver node, when receiving the message, 
verifies and 
checking the information and 
carrying out
the similarity calculation. 
Once 
the~\cp{similarity threshold ($CThresh$)} is met, the node joins the cluster, otherwise
it becomes part of the suspect list at first.
Fig.~\ref{Fig:funci} shows 
how CONFINIT acts over the cluster formation and leader election. Solid edges indicate nodes with transmission range overlap and that can exchange messages between them. The boxes next to each node mean, from top to bottom, the individual reading of the node,  neighbors nodes aggregate reading, and the value of aggregate readings.
Thus, we considered the 
($CThresh$) equal to three for cluster formation. Each instant~\textit{T} corresponds to a message exchange round among nodes in order to form clusters and elect leaders. 

In this way, assuming CONFINIT makes use of a ($CThresh$) =~\textit{3}, 
in $T_1$ all nodes start exchanging data messages to update the neighbors list and aggregate readings taking into account 
Equation~\ref{eq:firefc}. They begin the process of forming clusters and electing leaders based on the readings obtained at the earlier instant~$Ra_{Tn-1}$ according to Equation~\ref{eq:firefc}. 
In $T_2$, 
we have the following 
similarity between the nodes, $Ra_{T2}(n_a) = \frac{15+1*16}{1+1}$,
$Ra_{T2}(n_b) = \frac{16+1*15+1*18}{1+1+1}$,
$Ra_{T2}(n_c) = \frac{18+1*16+1*16+1*16+1*17}{1+1+1+1}$,
$Ra_{T2}(n_d) = \frac{17+1*16+1*18}{1+1+1}$,
$Ra_{T2}(n_e) = \frac{16+1*17+1*18}{1+1+1}$.
After that, node~\textbf{$n_b$} 
has the largest number of neighbors, being elected as the cluster leader. As the clustering process works dynamically on each node,  
in $T_3$, it is done again due to existence of a new exchanged data message among the nodes,  thus the new similarity calculation is applied to maintain the cluster formation,
$Ra_{T3}(n_a) = \frac{20+1*22+1*21+1*23}{1+1+1+1}$,
$Ra_{T3}(n_b) = \frac{22+1*20+1*23}{1+1+1}$,
$Ra_{T3}(n_c) = \frac{23+1*22+1*20}{1+1+1}$,
$Ra_{T3}(n_d) = \frac{21+1*24+1*20}{1+1+1}$,
$Ra_{T3}(n_e) = \frac{24+1*21}{1+1}$.
The clustering configuration enables nodes to know which neighbors belong to the same cluster and it ensures better network scalability. It also helps to classify nodes with divergent readings, thus facilitating the identification of attackers by the fault control module.

The fault detection works over the formation of clusters. Thus, nodes whose readings do not respect the similarity threshold will be 
firstly
added 
into a suspect list.
The classification of nodes that were not part of the cluster begins in the control message, which without properly filled fields are discarded. Over 
the cluster formation,
nodes physically close to its neighbors at a given time and with distinct readings outside of the similarity threshold can be  
considered a suspicious node at the first time. 
Thus it becomes part of the suspect list, which contains nodes that have behaved anomalously but are not necessarily attackers. 
When a given node, which belongs to the suspects list, tries to join the cluster, and it again fails due to its distinct readings, then the system classifies it as attacker.
Hence, the system adds its $Id$ into a list containing all $Ids$ of threats. The cluster leader alerts the others
by sending an alert message. Thus,~when the attacker tries to join the cluster again, it is blocked. 

\begin{figure}[ht]
\centering    
\includegraphics[width=1\linewidth]{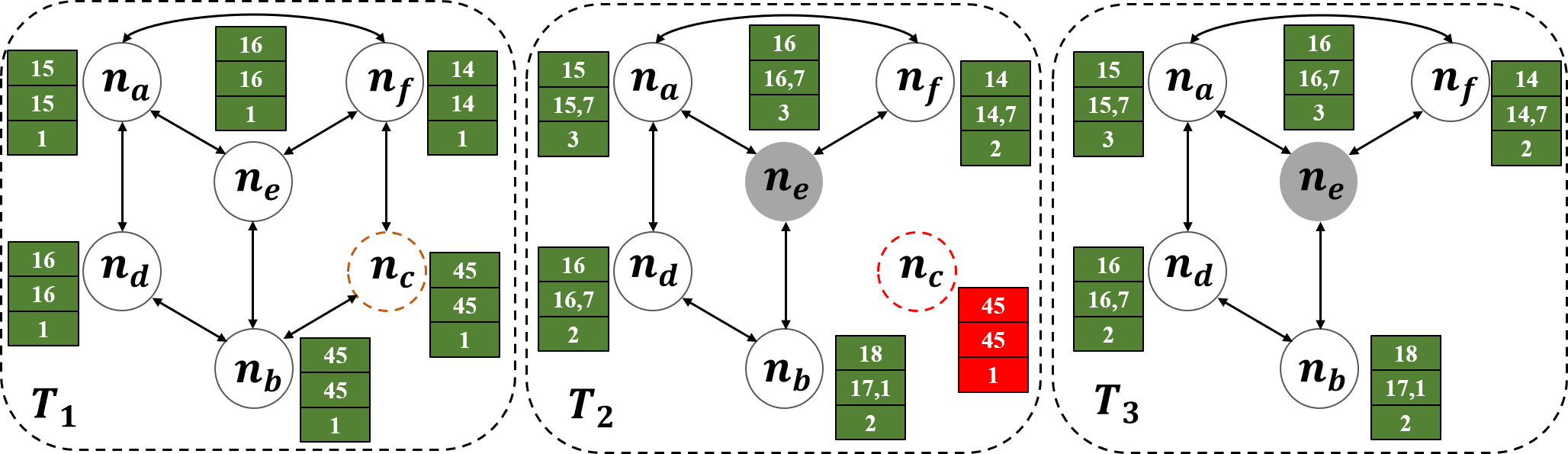}
\caption{Detection and isolation of an FDI Attack} 
\label{Fig:ata}
\end{figure}

Fig.~\ref{Fig:ata} illustrates 
an example of 
attack detection, where CONFINIT makes use of consensus threshold ($Thresholdconsensus$) =~\textit{5}.
Each instant $T$ means a clustering process and whose nodes that meet the ($CThresh$) 
value 
become clustered. Thus, at the instant $T_1$ nodes~\textbf {$(n_a, n_b, n_d, n_d, n_f)$} have 
reading values varying from $14$ to $18$, respecting the 
($CThresh$) among them. 
But, node~\textbf{$n_c$} has a value of $45$ for its reading, which deviates greatly from its spatial neighbors, not respecting the ($CThresh$). 
Hence, node~\textbf{$n_c$} 
cannot integrate the cluster at that time. 
At $T_2$, node~\textbf{$n_c$} again tries to join the cluster, but its Id is already on the suspect list. Thus, Equation~\ref{eq:conse}  applies to check the ($Thresholdconsensus$) based on consensus participates readings and compared with node~\textbf{$n_c$}. In this way, they determined~\textbf{$n_c$} is an attacker and cannot integrate any cluster.  At $T_3$, node~\textbf{$n_c$} is removed from the network, and an alarm message sent to the leader with the $Id$ of the attacker. After receiving the message, the leader disseminates the message to the other leaders on the network.

\begin{figure*}[ht]
    \centering   
   \includegraphics[width=0.30\linewidth]{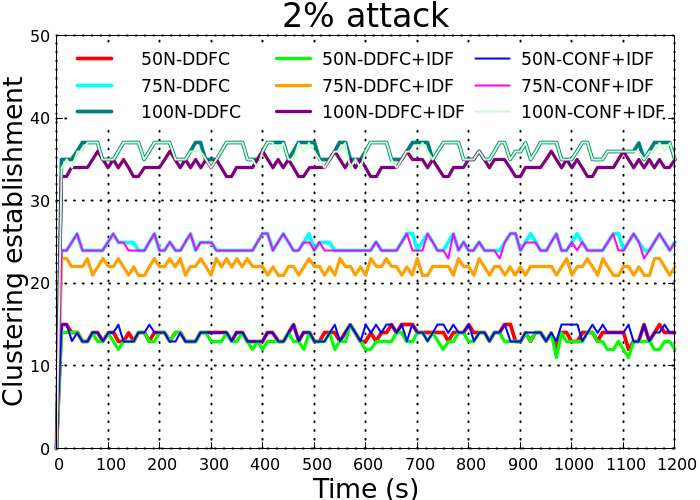}
   \includegraphics[width=0.30\linewidth]{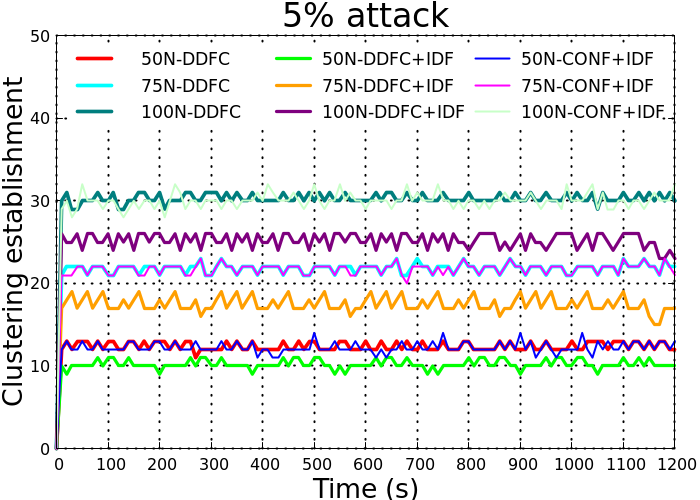}
    \includegraphics[width=0.30\linewidth]{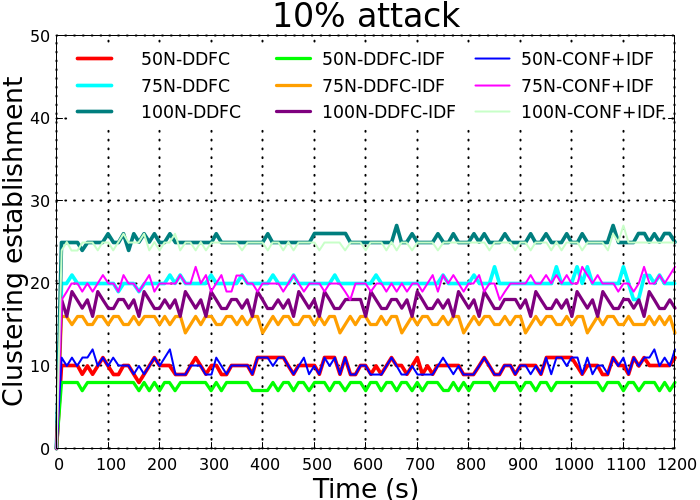}
      \caption{Number of clustering establishment over time}
      \label{Fig:fun}
\end{figure*}

\section{Analysis}
\label{sec:ana}
This section  shows a performance evaluation of CONFINIT. We implement CONFINIT in NS3-simulator, 
a data set of gas pressure sensors collected from a real petrochemical industry environment available by the UCI Machine Learning Repository lab~\cite{UCI} in order to get close to a real industrial environment. We also implemented the behavior of FDI attack based in~\cite{deng2016false}, whose attackers aim to alter data readings. We assume that the~attacker works synchronously and knows the type of the network traffic data,  facilitating the interaction for modifying the collected values, meaning the closest behavior to a FDI attack. The industrial scenarios consisted of $50$, $75$, and $100$ homogeneous nodes randomly distributed in a rectangular area of~\textit{200 x 200} meters operating by~\textit{1200s} with a transmission radius of~\textit{100m}. This transmission radius value is proportional to the coverage area in order to  not interfere with the clustering establishment and the collaborative consensus, and hence not compromising the achieved results. We defined the similarity threshold ($CThresh$) with the value $3$ and the consensus threshold ($Thresholdconsensus$) with the value $5$. We took these thresholds based on the type of data obtained by the~\textit{dataset}. These thresholds may vary according to the type of data and the application model. Moreover we do not consider energy issues once the nodes operate embedded in industrial equipment. The amount of FDI attacks ranges among $2\%$, $5\%$, and $10\%$, all nodes run IPv6 establishing an \textit{ad-hoc} standard IEEE 802.15.4. We also implement the DDFC~\cite{gielow2015data} clustering protocol as part of the CONFINIT system. Furthermore, the message controller, the adaptive interval, and the IPv6 integration from DDFC have been changed to better adapt it to the dense IIoT network context. The results obtained in all simulations correspond to the average values of 35 simulations with confidence interval of $95\%$. Table~\ref{tab:metricas} summarizes the metrics assessed according to~\cite{yang2017robust}.

\begin{table}[ht]
\renewcommand*{\arraystretch}{1.4}
\centering
\caption{Evaluation Metrics}
\label{tab:metricas}
{ 
\footnotesize
\begin{tabular}{|l|l|}
\hline
\textbf{METRIC
} & \textbf{EQUATION
} \\ \hline \hline

\textbf{\textbf{Detection rate (DR)}} 
& $({R_{det}}) = \frac{\sum {det _{ni}}}{{A _{ins}}}$ \\ \hline

\textbf{\textbf{Accuracy} (AC)}&$({A_{a}}) =\frac{\sum {det _{ni}} + \sum {det _{nl}}}{{R _{det}} + {det_{ni}} + {R_{fn}} + {R_{fp}}}$\\ \hline

\textbf{False positive rate (FPR)} & $({R_{fp}}) =\frac{\sum {det _{ni}}}{{T_{int}}}$ \\ \hline

\textbf{\textbf{False negative rate} (FNR)} & $({R _{fn}}) = |X| -{R_{det}}$ \\ \hline

\end{tabular}
}
\end{table}
\begin{figure*}[ht]
    \centering

    \includegraphics[width=0.23\linewidth]{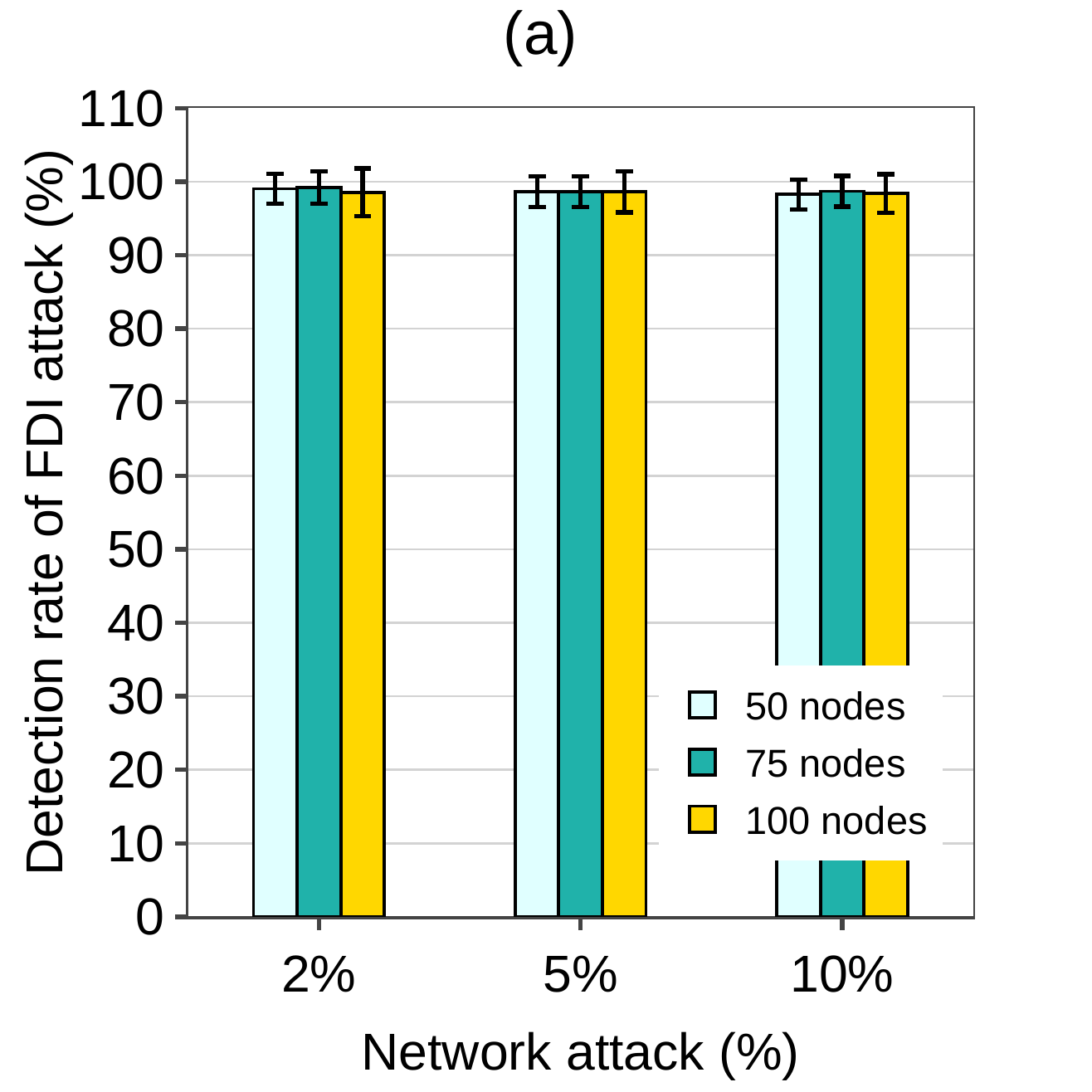}
    \includegraphics[width=0.23\linewidth]{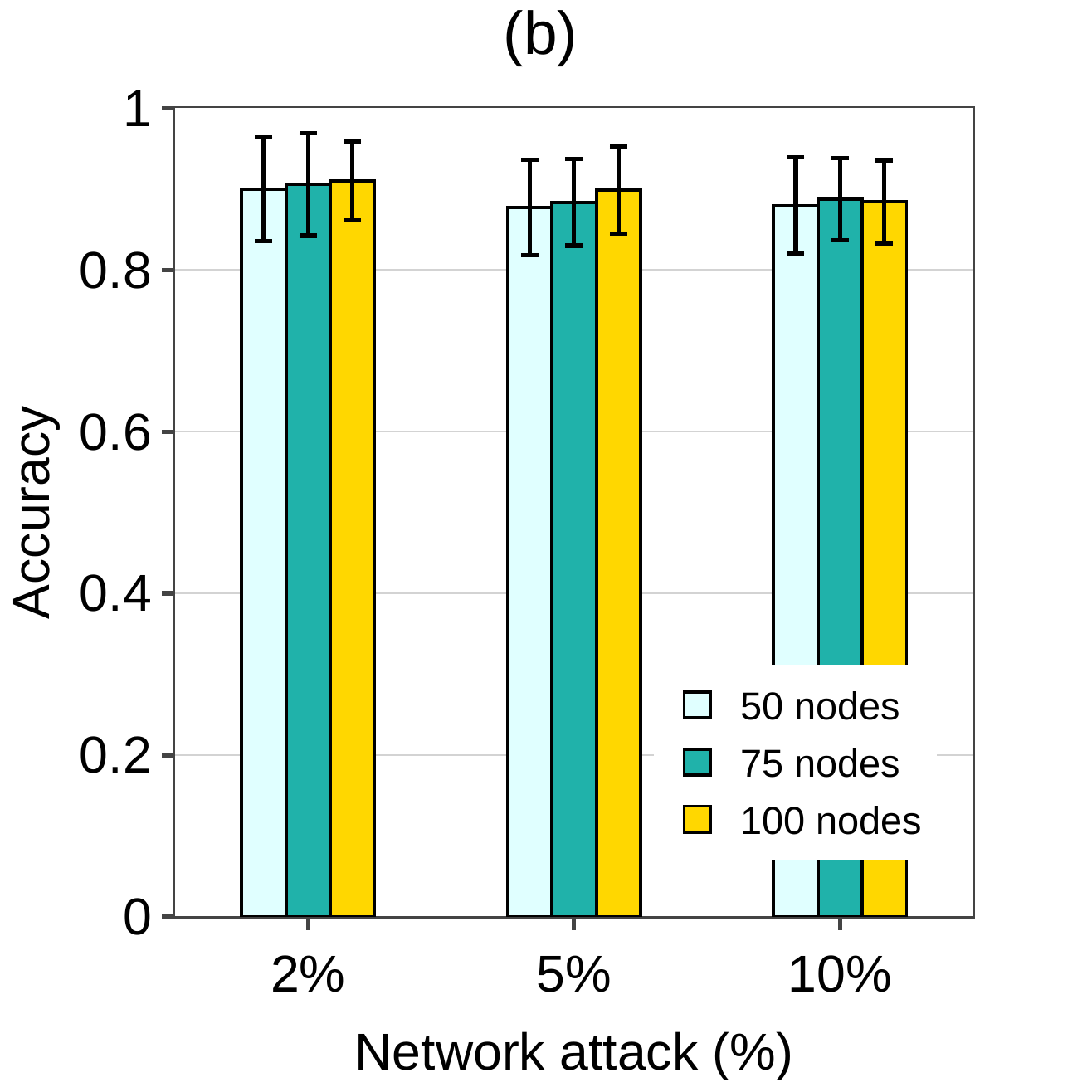}
    \includegraphics[width=0.23\linewidth]{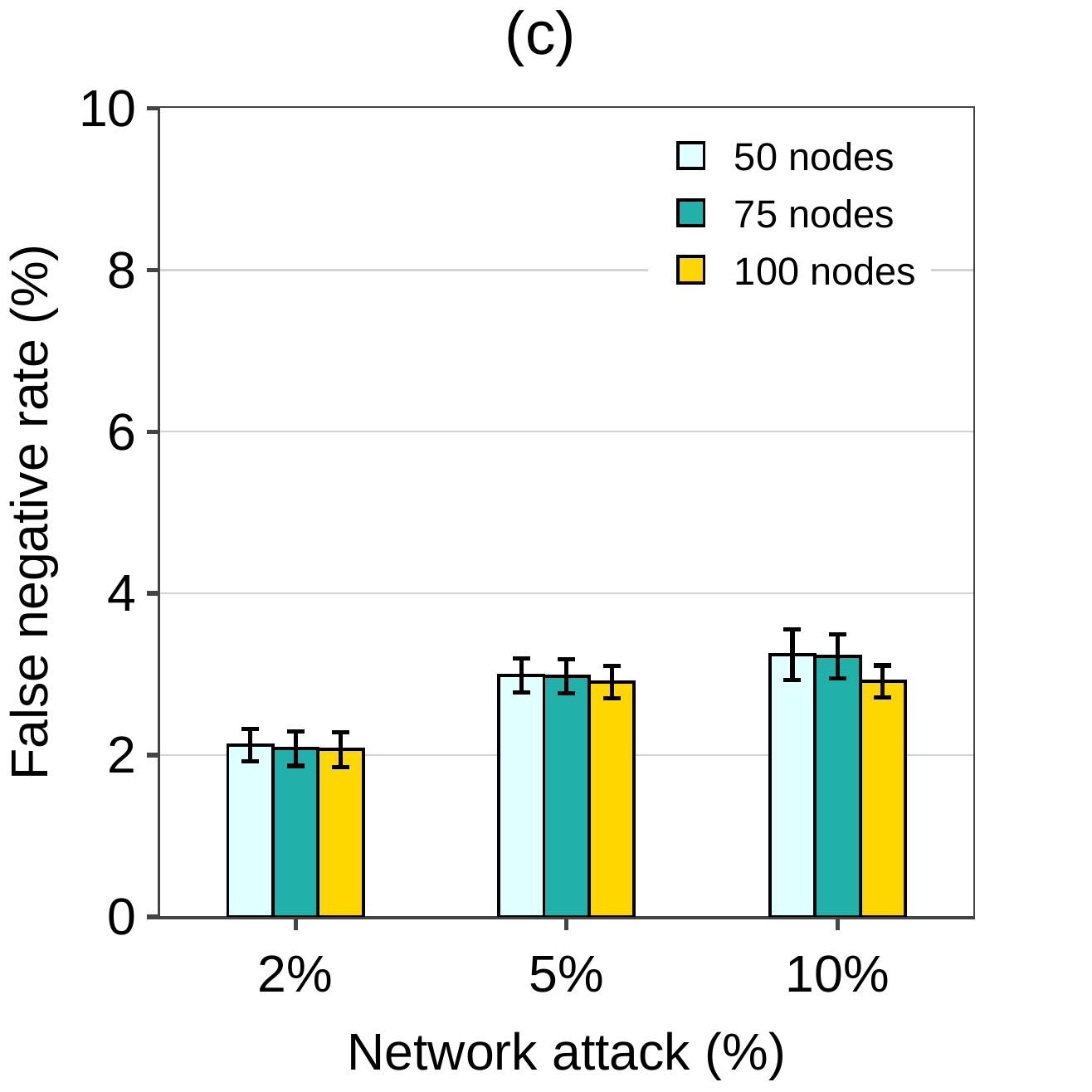}
    \includegraphics[width=0.23\linewidth]{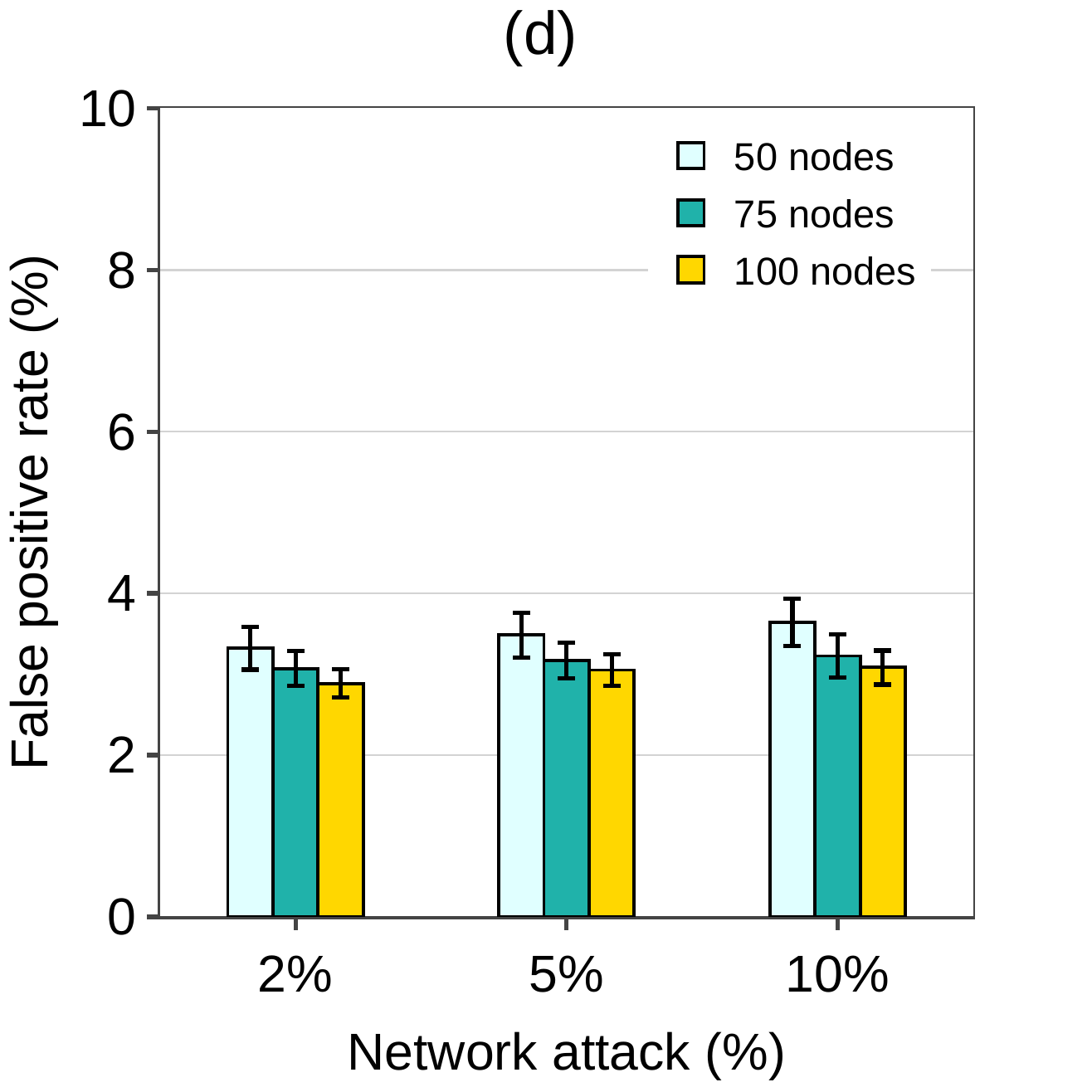}

      \caption{Detection rate of FDI attacks, 
      Accuracy, 
      False negative 
      and False positive} 
      \vspace{-0.2cm}
      \label{Fig:txdetec1}
\end{figure*}

\subsection{Clustering}
The CONFINIT performance for supporting the availability of the established clusters based on the number of nodes and the presence of attackers into the network is shown in Fig.~\ref{Fig:fun}. Note that the number of available clusters varies depending on the relationship of nodes and the amount of attackers present in the network. Regarding the FDI attack, it directly impacts on the number of clusters formed over the time.  Moreover, as expected, the scenario with $10\%$ 
of attacks had a stronger effect on the cluster establishment
than $5\%$ and $2\%$. In relation to DDFC, 
in some
cases, 
CONFINIT increases the number of clusters by $35\%$. This behavior has an effect on the amount of available data,  and 
interferes positively with the decision making. 
Such performance 
shows its efficiency in keeping the clusters safe and available. The formation of clusters considers the fact that the scenario is static, meaning, it  does not suffer changes in the node position; and only the readings vary in their values. 
Moreover, clusters hold a non-deterministic number of members, i.e., there is no fixed number of members.

  

\subsection{Attack detection} 
The detection and mitigation effectiveness of CONFINIT 
are shown in Fig.~\ref{Fig:txdetec1}~\textbf{(a)}.
It obtained an average DR of $97\%$, sometimes even reaching  $100\%$ depending on the variation of the number of nodes into the network. The CONFINIT accuracy is shown in Fig.~\ref{Fig:txdetec1}~\textbf{(b)}, whose values range from $0$ to $1$, being closer to $1$ more accurate. 
The AC values achieved by CONFINIT between $0.81$ and $0.98$ 
are due to the watchdog surveillance among participants, which evaluates the exchanged messages. 
Furthermore, the collaborative consensus building ensures better validation and high detection rate among nodes. 
As the scenarios are statics, we have also noted slight variations in the detection rate, showing that 
CONFINIT high ability to handle FDI attacks in dense environments. 

The impact of percentages 
of attacks on the false negative rates is 
shown in Fig.~\ref{Fig:txdetec1}~\textbf{(c)}.
The CONFINIT FPR varied from $2\%$, with $2\%$ of attackers, to $3.2\%$ with $10\%$ of attackers. The average variation 
among scenarios was only $1.2\%$, meaning CONFINIT
detects the most of attacking nodes. 
The attacker's detection failure emerged 
when there were errors in the similarity computation stage by neighbor nodes, and 
certain nodes became suspect node.
Thus, in the consensus stage, 
there were
a misidentification of these nodes as attackers. 
The CONFINIT FPR,  
Fig.~\ref{Fig:txdetec1}~\textbf{(d)}, ranged 
from $2.8\%$, with $2\%$ of attackers to $3.6\%$ with $10\%$ of attackers. 
The slight 
difference 
of $0.8\%$ for all scenarios shows a low variation among them. 
Detection failures 
were also due to errors in the consensus computation 
among monitor nodes of an attacker, which took 
a low deviation from their readings. Therefore, firstly nodes 
were 
added to 
the 
individual suspect list, but with 
new interactions and hence message exchanges, new 
computations pointed out such nodes as common nodes.

\section{Conclusion}
\label{sec:con}

This paper presented the CONFINIT system for mitigating false data injection attacks in dense IoT networks. CONFINIT arranges an IoT network in clusters  through the reading similarity among nodes to deal with the density issue. CONFINIT also embraces watchdog strategy and collaborative consensus to monitor misbehavior nodes concerning their read information, neighbors, and aggregate readings to know nodes with malicious behavior about to others. Results have shown the CONFINIT effectiveness against FDI attackers and insuring only the availability of legitimate data. As future work will be evaluated under other contexts of dense IoT that demand data clusters, and the impact of the energy consumption and mobility.

\bibliographystyle{IEEEtran}
\bibliography{sbc-template.bib}

\end{document}